# On the Automated Segmentation of Epicardial and Mediastinal Cardiac Adipose Tissues Using Classification Algorithms

Érick Oliveira Rodrigues[a], Felipe Fernandes Cordeiro de Morais[b], Aura Conci[a]

[a] *Institute of Computing, Universidade Federal Fluminense (UFF), Niterói, RJ, Brazil*
[b] *School of Medicine, Universidade Federal do Rio de Janeiro (UFRJ), Rio de Janeiro, RJ, Brazil*

## Abstract

*The quantification of fat depots on the surroundings of the heart is an accurate procedure for evaluating health risk factors correlated with several diseases. However, this type of evaluation is not widely employed in clinical practice due to the required human workload. This work proposes a novel technique for the automatic segmentation of cardiac fat pads. The technique is based on applying classification algorithms to the segmentation of cardiac CT images. Furthermore, we extensively evaluate the performance of several algorithms on this task and discuss which provided better predictive models. Experimental results have shown that the mean accuracy for the classification of epicardial and mediastinal fats has been 98.4% with a mean true positive rate of 96.2%. On average, the Dice similarity index, regarding the segmented patients and the ground truth, was equal to 96.8%. Therfore, our technique has achieved the most accurate results for the automatic segmentation of cardiac fats, to date.*

## Introduction

Extensive studies address the importance of the epicardial and mediastinal (due to a nomenclature inconsistency some call it pericardial) fats and their correlation with pathogenic profiles, risk factors and diseases [1,2]. Some work [3] associates both mediastinal and epicardial fats to carotid stiffness; others [4] associate them to atherosclerosis, coronary artery calcification and others health risk factors. Sicari *et al.* [5] have shown as well how mediastinal fat rather than epicardial fat is a cardiometabolic risk marker.

An increasing demand for medical diagnosis support systems has been observed jointly to the computational evolution in the last years. Systems of this kind speed up the tedious and meticulous analysis conducted by physicians or technicians on patients' medical data. In several cases there is a huge amount of data to be analyzed and, therefore, the diagnosis may lack of precision and suffer noticeable inter and intra-observer variation [6].

The automated quantitative analysis of epicardial and mediastinal fats certainly adds a prognostic value to cardiac CT trials with an improvement on its cost-effectiveness. Iacobellis *et al.* [7] have shown that the epicardial fat thickness and coronary artery disease correlate independently to obesity, which supports individual segmentation of these adipose tissues rather than merely and simply estimating its volume based on the patient overall fat.

Latterly, three imaging techniques appear suitable for quantification of these adipose tissues, namely Magnetic Resonance Imaging (MRI), Echocardiography and Computed Tomography (CT). Each has been used in several medical studies [5,8,9]. However, CT provides a more accurate evaluation of fat tissues due to its higher spatial resolution compared to ultrasound and MRI [10]. In addition, CT is also widely used for evaluating coronary calcium score [9].

## Literature Review

Semi-automated segmentation methods for epicardial fat have been proposed since 2008. Dey *et al.* [11], for instance, apply a preprocessing step to remove all other structures from the heart by using a region growing strategy. An experienced user is required to scroll through the slices and to place from 5 to 7 control points along the pericardium border if visible. Thus, Catmull-Rom cubic spline functions are automatically generated to obtain a smooth closed pericardial contour. Finally, since the epicardial fat is inside this contour it is simply accounted by thresholding. In [12], a method is proposed for the segmentation of abdominal adipose tissue. The work of Kakadiaris *et al.* [13] has further extended the method to the segmentation of the epicardial fat.

Coppini *et al.* [10] focused on reducing the user intervention. On their method, an expert is still necessary to scroll through the slices between the atrioventricular sulcus and the apex in order to place some control points on the pericardium layer. The amount of required control points is not clearly described. Nevertheless, the amount of slices to be analyzed is apparently less than that of the method proposed by Dey *et al.* [11]. Coppini et al. also present their solution in a 3D space. The overall focus of their work was to describe their method mathematically. Nevertheless, the authors did not address the achieved general accuracy.

Barbosa *et al.* [14] proposed a more automated segmentation method for the epicardial fat. They use the same preprocessing method from Dey *et al.* [11] and further apply a high level step for identification of the pericardium by tracing lines originating from the heart's centroid to the pericardium layer and interpolating them with a spline. Although this approach may be of simple complexity and highly applicable for virtually any proposed method in the field, the reported results are not impressive. Only 4 out of 40 images were correctly segmented in a fully automatic way.

As far as we know, Shahzad *et al.* [15] proposed the first fully automated method for epicardial fat segmentation. Their method uses a multi-atlas based approach to segment the pericardium. The multi-atlas approach is based on registering several atlases (8 in this case) to a target patient and fusing these transformations to obtain the final result. They selected 98 patients for evaluation and reported a Dice similarity index of 89.15% to the ground truth and a low rate of approximately 3% of unsuccessful segmentations. Nevertheless, they did not provide any measurements of the overall processing time.

Ding *et al.* [16] proposed, in 2010, an approach similar to the method of Shahzad *et al.* [15]. They segment the pericardium using an atlas approach, which consists of a minimization of errors after applying transformations to the atlas along with an active contour approach. The mean Dice similarity coefficient was 93% and they claim that their result was achieved in 60 seconds on a simple personal computer. Although their segmentation seems to be the most precise in the literature, the reported computational time is oddly described. That is, 60 seconds is considered too fast for segmenting and transforming an entire patient scan, which consists of roughly 50 images. They also present a work [17] that segmented the aorta instead of the pericardium and compare their time (60 seconds) to the 15 minutes of the former. If these 60 seconds correspond to just the time it takes for the algorithm to minimize the transformations, then this comparison is not feasible. Furthermore, they report that their approach had the atlases' images pre-aligned to a standard orientation and that there is a comparison with just one of these atlases to speed up the process. The remaining pericardium contour follows the pre-aligned pattern, which is a reported limitation. Besides, they did not describe how each one of these atlases is chosen as suitable for each possible case.

## Materials and Methods

We define the fat located within the epicardium as epicardial fat, corroborating with the majority of published works [4-8,10]. Furthermore, by following the same "first outer anatomical container" logic, we conclude that mediastinal fat is the best definition for the fat located on the external surface of the heart or fibrous pericardium. In other words, the mediastinal fat is located within the mediastinal space as long as it is not epicardial (i.e., it is not located within the epicardium). Furthermore, we have used CT scans from two manufactures (Siemens and Philips), which configures this work as being multi-manufacturer.

Molteni [18] associates values around -100 Hounsfield Units (HU) to the overall fat of the human body. Coppini *et al.* [10] and Shmilovich et al. [19] defined the cardiac adipose tissue range as starting from -190 to -30, Spearman *et al.* [20] defined it as from -195 up to -45, and Shahzad *et al.* [15] defined it as from -200 to -30. In this work, we consider the largest addressed interval for the cardiac fat, which corresponds to the one considered in the work of Shahzad *et al.*, *i*.e., from -200 to -30 HU.

### Overview

We propose an automatic segmentation for the epicardial and mediastinal fats based on two main principles, namely (1) an intersubject registration and (2) a classification step. For the whole segmentation process we have used CT images ranging from -200 to -30 HU on the axial plane. However, we believe that our methodology can be easily adapted to other ranges and also to other modalities.

Image registration can be defined as the process of matching characteristics of images in order to find alignments that minimize the variation between overlapping pixels [22]. Such processes are included in panoramic assemblages, medical images, time series alignments [23,24] and many others. Registration is also alternatively treated as an optimization problem with the goal of finding the spatial mapping that will bring images, parts of them, or even a combination of these parts into minimal variation.

Machine learning algorithms are often divided in two main categories: (1) supervised and (2) unsupervised methods. The algorithm is categorized as supervised when it explicitly evaluates the class attribute of a training set as the predictive label desired to attach to an incoming unlabeled instance. Furthermore, when this assumption is formalized, the class attribute heavily induces the generated predictive model. However, without the formalization, the algorithm is defined as unsupervised and the class induces no heavy influence but of a normal attribute to the predictive model when, of course, it is not disregarded from the training pace. Classification algorithms are always categorized as supervised learning methods while clustering algorithms are often unsupervised.

### Registration

In a previous work we extensively described the methodology of our intersubject registration [25]. Summing up, it is composed of (1) creating an atlas of the retrosternal area, (2) using this atlas to recognize the same area of an incoming instance, and (3) using a heuristical method to reinforce the chosen position. The atlas assemblage is done by (1) converting a number of CT slices to the fat range where black represents the background, (2) manually selecting the position of the retrosternal area, (3) thresholding each selected area to 0 or 1 (binary image) and finally, (4) performing an arithmetic mean of the binary images. The final image of Figure 1 is the result of the described process conducted for 10 randomly chosen patients. Thereafter, the proposed registration consists of fusing two approaches: a landmark with an atlas approach. The landmark approach is defined as when two selected points from the subjects to be registered are used as reference (in this case, the center of the retrosternal area) [26]. Finally, the retrosternal area is defined as the region at the back of the sternum and, in this case, it frequently captures parts of the heart if it is relatively close to the thorax as well as few parts of the thorax as shown in Figure 1 by the rectangles in red.

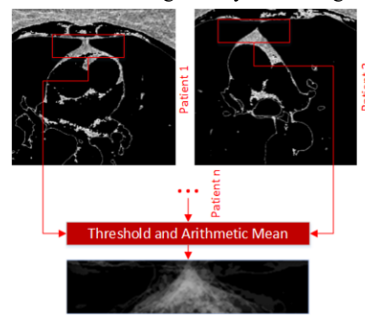

*Figure 1 – Atlas of the retrosternal area.*

Once the atlas is assembled it is considered immutable for any subject to be registered. The following step is to displace the atlas image on top of any cardiac CT image that is on the proper fat range and that has the retrosternal area visible and, to compute a similarity score associated to every possible position of the atlas. A weighted mutual information (WMI), as shown in Equation (1), was chosen for computing the similarity score. Hence, we are autonomously recognizing the pattern of the retrosternal area on an arbitrary patient in order to suppose its location. After assuming its location on the basis of the WMI score, a heuristical confirmation method was applied to reinforce the chosen position. Once the position is set, all the retrosternal areas of the patients being registered are aligned, based on the recognized landmark, to a common position. It is important to emphasize that this registration step is required to be done on a single slice of each patient; and the same transformation is applied to the remaining.

$$WMI_{y,x}(F,M,g) = \left(\sum_{f \in F} \sum_{m \in M} \frac{1}{|f-m|+1} \rho_{FM}(f,m) \log_g \frac{\rho_{FM}(f,m)}{\rho_F(f)\rho_M(m)}\right) \quad (1)$$

The variable $F$ represents the fixed slice, $M$ represents the moving atlas and $g$ stands for the base of the logarithm. In the traditional formulation of the mutual information, each event or object specified by $(f,m)$ is weighted by the corresponding joint probability $\rho_{FM}(f,m)$. This assumes that all objects or events are equivalent apart from their probability of occurrence. However, in some applications, it may be the case that certain objects or events are more significant than others, or that certain types of associations are more semantically important than others. Thus, we combined the mean difference measure with the mutual information, originating a weighted measure by the difference mutual information measure previously shown in Equation (1).

**Classification**

The classified segmentation can be viewed as a simple iteration through a set of pixels of an image (or voxels of a 3D model) where a set of features (i.e., characteristics) related to the iterated pixel, voxel, or surrounding area is extracted. The set of features is commonly called the features vector. The vectors are often united to compose a dataset that is provided as input to a classification algorithm.

In order to generate a concise predictive model we need to provide reliable data for training the classification algorithm. Therefore, two specialists, one being a physician and the other being a computer scientist, have manually segmented the epicardial and mediastinal adipose tissues of 20 patients (10 male and 10 female). Thus, our ground truth contains approximately 900 manually segmented cardiac CT images. It is important to highlight that prior to the manual segmentation, the images were already registered by our methodology. The created ground truth is available at [27]. The black value (0) represents the background and these pixels are excluded from the feature extraction and from the statistics. All other pixel values represent the selected cardiac fat range: (-200,-30) HU.

Similarly to the approach of Rikxoort *et al.* [28] for the classified segmentation of the liver, we have selected features: the pixel grey level and the position x, y and z, where z is the index of the slice. Besides, we have also selected the x and y positions relative to the center of gravity of the image and texture-based features from a neighborhood of variable size that encapsulates the iterated pixel at its center, i.e., a surrounding window of pixel values. Some features were selected on a theoretical basis and extracted from this neighborhood, such as: (1) a simple arithmetic mean of the grey levels, (2) moments of the co-occurrence matrix, (3) geometric moments, (4) run percentage, (5) grey level non-uniformity and (6) a 1D Gaussian-weighted mean of the grey values [24]. There are three possible classes for a pixel in our problem, namely, (1) epicardial fat, (2) mediastinal fat or (3) pixel of the pericardium (or unknown). These classes were evaluated separately and, therefore, the problem was reduced to a binary classification (true or false) for each one of these three possible classes [24]. If an incoming pixel is classified as epicardial and mediastinal, the result is a hybrid class of both. If it is classified as pericardium and has no current label, the result is also the same hybrid (further represented in yellow in Figure 3).

**Results**

Following feature selection, some classifiers were selected for a first evaluation of the convergence speed. For this trial, we extracted the features from the slices of just one single patient (approximately 13 107 200 features vectors) in order to speed up the analysis. All the classification and clustering algorithms in Weka [30] were selected for evaluation, some of these algorithms are, namely, the Support Vector Machine (SVM), Sequential Minimal Optimization (SMO), Naïve Bayes, Radial Basis Function (RBF) Network, Random Trees, C4.5 (or J48), Primal Estimated Sub-Gradient Solver for SVM (SPegasos), REPTree, k-Nearest Neighboor (kNN), Multilayer Perceptron and others. The parameters used by the evaluated classifiers were based on their standards with a small tweaking in some cases and the best result was selected. The algorithms present in Table 1 were the only ones that converged within an interval of 200 seconds (constructing and evaluating the predictive model). The values stand for mean accuracies and times of the three possible classes on a neighborhood size of 5x5 pixels.

*Table 1 – Accuracies and convergence time on a single patient*

| Algorithm | Accuracy | Time (s) | Acc/Time |
|---|---|---|---|
| J48Graft | 99.0% | 132.86 | 0.75 |
| RandomForest | 98.9% | 112.57 | 0.88 |
| REPTree | 98.9% | 10.34 | 9.56 |
| J48 | 98.9% | 151.23 | 0.65 |
| SimpleCart | 98.9% | 108.78 | 0.91 |
| SMO | 98.3% | 58.66 | 1.68 |
| RandomTree | 97.5% | 8.0 | 12.19 |
| RBFNetwork | 96.8% | 3.48 | 27.82 |
| SPegasos | 96.8% | 15.77 | 6.14 |
| DecisionStump | 96.8% | 52.34 | 1.85 |
| HyperPipes | 94.8% | 0.04 | 2370.0 |
| NaiveBayes | 86.0% | 55.48 | 1.55 |

At first glance, the REPTree algorithm appears to be the best choice since it achieved a great accuracy in a relatively short time, as compared to the others. Also of high interest was the HyperPipes result. In this case, it returned a good accuracy almost instantaneously. However, when applied to a bigger dataset and a set of images of distinct patients, the results of the HyperPipes algorithm happen to differ drastically.

**Overall Evaluation**

To avoid unfair comparisons of classifiers we have further evaluated some of the algorithms of Table 1 along with a variation on the neighborhood size (modified during the extraction of features). Hypothetically, one may consider that some classifiers perform better on a neighborhood of a certain specific size. In Figure 2, a chart that represents the accuracy (vertical axis) of each evaluated classifier (horizontal axis) versus the variation of the neighborhood size in pixels (orthogonal to the previous axes) is shown. The assessed CT images were 512x512 pixels wide.

The accuracies in Figure 2 were achieved using 66% random-selected split method as test mode from the data of 16 patients. The 66% split test mode is defined as randomly selecting 66% of the instances for training and using the remaining for evaluation of the predictive model. The reason for choosing the split method and 16 patients was to reduce the huge convergence time due to the great amount of data and, consequencely, to speed up the analysis. The extracted features composed datasets are approximately 1.5 gigabytes for each neighborhood size that was provided to each classifier. The period to train and evaluate the predictive model was, in some cases, up to 20 minutes for each possible combination of neighborhood size and classification algorithm.

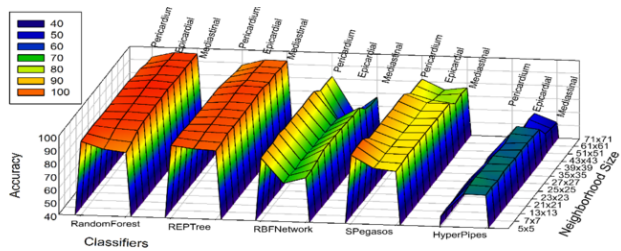

*Figure 2 – Accuracies of selected classifiers versus pixels neighborhoods of distinct sizes.*

The five algorithms shown in Figure 2 were the quickest on this large dataset (16 patients). The HyperPipes was the fastest and always converged virtually within 0.5 seconds but, in this case, the best accuracy that it could achieve was around 70%. We can state by this evidence that this algorithm is not generalizable. In other words, the algorithm overfitted to a single patient and failed to generalize the predictive model to a class instead of a single patient. It is a rather simple algorithm and, in fact, the achieved low accuracy was somehow supposed to happen, even in the convergence evaluation trial.

Due to the convergence time issue, we were not able to extensively assess all the possible sizes for the neighborhood in consideration for all the algorithms shown in Table 1. The REPTree algorithm did not converge remarkably faster than the remaining decision tree algorithms evaluated on this large dataset. The RBFNetwork was faster than RandomForest and SPegasos was slower but both achieved lower accuracies. SPegasos was the slowest among the algorithms in Figure 2. RandomTree and DecisionStump were just a little faster than RandomForest and REPTree but the accuracies were significantly lower; therefore, they were disregarded in the second evaluation due to the massive presence of decision tree algorithms. The J48Graft returned similar accuracies if compared to RandomForest but its convergence was approximately 1.4 times slower and, so it was impracticable to evaluate all the neighborhood sizes for this classifier. The SimpleCart, Naïve Bayes, J48 and SMO took more time to converge on the large dataset than SPegasos, therefore, they could not be precisely evaluated in the second trial.

**Visualization**

Figure 3 compares a single manually segmented slice (left) to the result of the proposed automatic segmentation (right). Green denotes the mediastinal fat, red represents the epicardial and blue corresponds to the pericardium. All the colored pixels represent pixels within the fat range for a CT image, therefore, there are some discontinuities on the images. Figure 4 corresponds to the same patient shown in Figure 3 but reconstructed in a 3D model. It is possible to distinguish the contour of the heart based on the epicardial fat (red color).

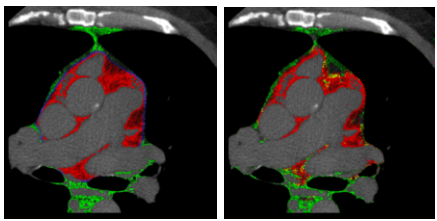

*Figure 3 – Manually (left) and automatically segmented slice (right).*

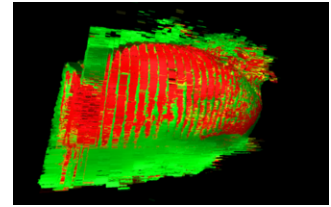

*Figure 4 – 3D model of the automatically segmented patient.*

Table 2 compares the results obtained by the proposed methodology to the four main related works. Barbosa *et al.* [14] and Kakadiaris *et al.* [13] use semi-automated methods, while Shahzad *et al.* [15] and Ding *et al.* [16] use fully automatic methods. The first column of the table indicates the rate of successful automatic segmentations, which are usually observed by a physician. The second column is the Dice similarity index and the third is the true positive rate. All these four works only segment the epicardial fat, therefore, we only compare our epicardial fat segmentation rates.

*Table 2 – Comparison of results*

| Authors | Successful | Dice | T.P. |
| --- | --- | --- | --- |
| Barbosa *et al.* | 10% (4/40) | - | - |
| Kakadiaris *et al.* | - | - | 85.6% |
| Shahzad *et al.* | 96.9% (95/98) | 89.15% | - |
| Ding *et al.* | - | 93.0% | - |
| This work (epicardial) | 100% (82/82) | 97.9% | 98.3% |

**Conclusion**

The proposed automated method achieved the best results with no need for placement of any control point, the mean accuracy for both the epicardial and mediastinal fats was 98.4% with a mean true positive rate of 96.2%. We are the first to propose an automated segmentation of the mediastinal fat and a unified methodology for the automated segmentation of both types of fat. Despite the processing time issues of our approach, this is still feasible for real time segmentation if properly adjusted. Although it requires approximately one day to segment a single patient, some heuristics could be used to speed up the classification step while an extensive selection and evaluation of features could grant an overall speed improvement. The speed up gain is also strongly related to the meticulousness of the approach. If a faster segmentation is desired, a per-area classification instead of a per-pixel classification may be applied and it should return worse but still adequate results.

Ensemble methods use multiple learning algorithms to obtain better predictive performance. In this work, we have selected a single classification algorithm (Random Forest) to segment the patients. As a further improvement, the RandomForest could be combined to the J48Graft and, perhaps, to the REPTree algorithms to increase the accuracy of the predictive model. These three algorithms, RandomForest, J48Graft, and REPTree, were the best in our performance analysis.

**Acknowledgements**

E. O. Rodrigues wants to thank CAPES. A. Conci wants to thank the CNPq project 302298/2012-6. F. Morais wants to thank Ilan Gottlieb.

## References

[1] Ahmad I, Hua B, Sockolow JA, et al. Correlation of Pericardial and Mediastinal Fat With Coronary Artery Disease, Metabolic Syndrome, and Cardiac Risk Factors. *Journal of Cardiovascular Magnetic Resonance*, 2009, 53(10): A283.

[2] Perseghin G, Molon G, Canali G, et al., Nonalcoholic Fatty Liver Disease Is Associated With Left Ventricular Diastolic Dysfunction in Patients With Type 2 Diabetes. *Cardiovascular and Metabolic Risk*. 2013, 58: S542-S543.

[3] Brinkley TE, Hsu FC, Ding J, et al, Pericardial fat is associated with carotid stiffness in the Multi-Ethnic Study of Atherosclerosis. *NMCD. Nutrition Metabolism and Cardiovascular Diseases*. 2011, 21(5): 332-338.

[4] de Vos AM, Stella PR, Gorter PM, et al. Relation of Epicardial and Pericoronary Fat to Coronary Atherosclerosis and Coronary Artery Calcium in Patients Undergoing Coronary Angiography. *American journal of cardiology*, 2008,102(4):380-385.

[5] Sicari R, Maria A, Petz R, et al. Pericardial Rather Than Epicardial Fat is a Cardiometabolic Risk Marker: An MRI vs Echo Study. *Journal of the American Society of Echocardiography*, 2011, 24, 1156-1162.

[6] Aspen Publishers Inc. Web-based diagnosis support system seeks to reduce delays in medical diagnoses. *Managed care outlook*. 2010, 23(13): 8.

[7] Iacobellis G, Lonn E, Lamy A, et al. Epicardial fat thickness and coronary artery disease correlate independently of obesity. *International Journal of Cardiology*. 2011, 452-454.

[8] Iacobellis G, Ribaudo MC, Assael F, et al. Echocardiographic epicardial adipose tissue is related to anthropometric and clinical parameters of metabolic syndrome: A new indicator of cardiovascular risk. *The Journal of clinical endocrinology and metabolism*. 2003, 88:5163-5168.

[9] McClelland RL, Jorgensen NW, Bild DE, et al. Coronary Artery Calcium Score and Risk Classification for Coronary Heart Disease Prediction. *Journal of the American Medical Association*. 2010, 303(16): 1610-1616.

[10] Coppini G, Favilla R, Marraccini P. Quantification of Epicardial Fat by Cardiac CT Imaging. *The Open Medical Informatics Journal*. 2010, 4: 126-135.

[11] Suzuki Y, Suzuki S, Dey D, et al. Automated quantitation of pericardiac fat from noncontrast CT. *Investigative Radiology*, 2008, 43(2): 145-153.

[12] Pednekar A. Automatic Segmentation of Abdominal Fat from CT Data. *Application of Computer Vision*, 2005, 308-315.

[13] Kakadiaris I, Bandekar A, Mao SS, et al. Automated pericardial fat quantification in CT data. *Journal of the American College of Cardiology*, 2006, 47 (4): 264A.

[14] Barbosa JG, Figueiredo B, Bittencourt N, et al. Towards automatic quantification of the epicardial fat in non-contrasted CT images. *Computer Methods in Biomechanics and Biomedical Engineering*, 2011, 14(10): 905-914.

[15] Shahzad R, Bos D, Rossi A, et al. Automatic quantification of epicardial fat volume on non-enhanced cardiac CT scans using a multi-atlas segmentation approach. *Medical Physics*, 2013, 40(9).

[16] Ding X, Terzopoulos D, Piotr J, et al. Automated Epicardial Fat Volume Quantification from Non-Contrast CT. *Medical Imaging*, 2014, 9034.

[17] Isgun I. Multi-Atlas-Based Segmentation With Local Decision Fusion—Application to Cardiac and Aortic Segmentation in CT Scans. *IEEE Transactions on Medical Imaging*, 2009, 28(7): 1000-1010.

[18] Molteni R. Prospects and challenges of rendering tissue density in Hounsfield units for cone beam computed tomography. 2013, 116 (1): 105-119.

[19] Dey D, Shmilovich H, Slomka PJ, et al. Threshold for the upper normal limit of indexed epicardial fat volume: Derivation in a healthy population and validation in an outcome-based study. *The American Journal of Cardiology*, 2011, 108(11): 1680-1685.

[20] Spearman JV, Costello P, Geyer LL, et al. Automated Quantification of Epicardial Adipose Tissue Using CT Angiography: Evaluation of a Prototype Software. *European Radiology*, 2014, 24(2): 519-526.

[21] Huang J, Hsieh MH, Chen YJ, et al. Extremely high coronary artery calcium score is associated with a high cancer incidence. *International Journal of Cardiology*, 2012, 155: 474-475.

[22] Ahmad I, Hua B, Sockolow JA. et al. Correlation of Pericardial and Mediastinal Fat With Coronary Artery Disease, Metabolic Syndrome, and Cardiac Risk Factors. *Journal of Cardiovascular Magnetic Resonance*, 2009, 53(10): A283-A283.

[23] Gerseghin G, Molon G, Canali G, et al. Nonalcoholic Fatty Liver Disease Is Associated With Left Ventricular Diastolic Dysfunction in Patients With Type 2 Diabetes. *Cardiovascular and Metabolic Risk*, 2013, 58: S542-S543.

[24] Rodrigues ÉO. Automated segmentation of epicardial and mediastinal fats using intersubject registration and classification algorithms. Thesis, Computer Science Institute, UFF, Niterói, RJ, 2015.

[25] Viergever MA, Maintz JBA. A survey of medical image registration. *Medical Image Analysis*, 1998, 2(1): 1-36.

[26] Pocock AC. *Feature Selection Via Joint Likelihood*. Manchester, 2012.

[27] Groundtruth for epicardial and mediastinal fats. *Visual Lab*, 2014. http://visual.ic.uff.br/en/cardio/ctfat/

[28] Rikxoort E, Arzhaeva Y, Ginneken B. Automatic segmentation of the liver in computed tomography scans with voxel classification and atlas matching. *3D Segmentation in The Clinic: A Grand Challenge*, 2007, 101-108.

[29] Rodrigues ÉO, Conci A, Borchartt TB, et al. Comparing results of thermographic images based diagnosis for breast diseases. Proceddings of Systems, Signals and Image Processing (IWSSIP), 2014, 39-42.

[30] Frank E, G. Holmes G, M. Hall et al. The WEKA Data Mining Software: An Update. *SIGKDD Explorations*, 2009, 11(1): 10-18.

**Corresponding Author**

Name and contact information of corresponding author.